# Evaluating the Performance of LLMs in Drafting NIH Data Management Plans

**Authors**: Nahid Zeinali[1], James O'Neil[1], Xuebin Dong[1], Rebecca Hofstein Grady[2], Maria Praetzellis[2], Brian Riley[2], John Chodacki[2], Bhavesh Patel[*,1]

**Affiliations**
[1] FAIR Data Innovations Hub, California Medical Innovations Institute, San Diego, California, United States of America, 92121
[2] California Digital Library, University of California Office of the President, Oakland, CA, United States of America, 94607

**Running Title:** Performance of LLMs in Creating NIH DMPs

*Corresponding author
Bhavesh Patel
FAIR Data Innovations Hub
California Medical Innovations Institute
11107 Roselle St.
San Diego, CA 92121
T: (510) 604-2815, Email: bpatel@calmi2.org

## Abstract

Sharing research data is essential to promote transparency, enable reproducibility, and fuel scientific discoveries. To ensure consistent sharing, the National Institutes of Health (NIH) and other funding agencies require applicants to submit data management plans (DMPs) with their proposals. They aim to encourage researchers to consider their data management and sharing strategies early on. However, researchers often face challenges when developing DMPs due to a lack of knowledge about data-sharing practices. Here, we evaluated the performance of Large Language Models (LLMs) in drafting DMPs compliant with the 2023 NIH guidelines. The goal was to investigate whether LLM-supported DMP creation could reduce the burden on researchers and librarians. We compared Llama and GPT performance against NIH-provided DMP samples. We also asked domain experts to evaluate LLM-generated DMPs. Overall, we found that LLMs can accelerate the creation of initial drafts but still require expert review before submission to ensure clarity and completeness. Moreover, open-source LLMs may not be as ready off-the-shelf as commercial ones for drafting funder-compliant DMPs.

**Keywords**: Data Management Plan, Data Sharing, Standards, Funding, Grant, Large Language Model, Prompt Engineering

## Introduction

Research data are fundamental to scientific progress.[1] It is estimated that data generated by other researchers serve as the primary source for more than half of published scientific manuscripts.[2] Therefore, data sharing is essential not only for ensuring transparency and enabling reproducibility but also for accelerating discoveries.[3] The scientific community has undertaken substantial efforts to standardize and promote data sharing. Most notably, the Findable, Accessible, Interoperable, and Reusable (FAIR) principles provide high-level guidelines for improving the discovery, access, interoperability, and reuse of research data.[4] Since their introduction, the FAIR principles have been broadly adopted by funding agencies, publishers, research institutions, and data repositories worldwide.[5] Despite the broad recognition of the FAIR principles and their benefits in the scientific community, data sharing is still not a consistent practice.[6]

To address this, funding agencies have begun requiring Data Management Plans (DMPs, sometimes also called Data Management and Sharing Plans or DMSPs) with grant proposals. In their DMPs, grant applicants are typically required to describe how they will manage, preserve, and share project data in line with FAIR principles and other relevant disciplinary guidelines.[7] The National Institutes of Health (NIH) initially introduced in 2003 a data sharing policy that required a brief DMP for projects requesting $500k or more in direct costs per year.[8] This policy was substantially expanded in January 2023, requiring all grant applicants to include a DMP with their proposal, irrespective of the amount requested.[9] Similarly, the National Science Foundation (NSF) mandated DMPs for all proposals submitted starting January 2011.[10] With DMPs, funding agencies aim to encourage researchers to consider their data management and sharing strategies, as well as FAIR practices, even before their projects start. They also use DMPs to ensure that grantees comply with their data-sharing plans.[1] Overall, the goal of these efforts is to promote long-term preservation of digital scientific data while supporting ethical, technical, and logistical planning for data reuse.[11]

Writing DMPs, however, can be challenging for researchers because they often lack training in FAIR data practices.[12] Additionally, DMPs from different funding agencies differ substantially in terminology, formats, and requirements, accentuating challenges for researchers.[13] Librarians are typically available at larger universities to help researchers with their DMPs, but researchers at independent and smaller institutions lack such support.[14] This also introduces a burden on librarians, who may have varying levels of training in FAIR data practices and awareness of funder-specific requirements.[15] Open source and free platforms, such as DMP Tool, are available to support researchers by providing funder-specific guidance, sample DMPs, and a convenient user interface for creating compliant DMPs.[16] Yet, it is still up to the researchers and librarians to fill out the different sections of their DMPs, which vary greatly based on many factors such as research domain, sensitivity of data, and funder requirements.[17]

Recent advances in Large Language Models (LLMs) have demonstrated their potential to assist with complex technical writing tasks, including drafting patent abstracts and clinical trial protocols.[18,19] Given the structured nature of DMPs, LLMs represent a promising tool to support researchers and librarians, but have not yet been evaluated for this task. As an initial step toward understanding the broader potential of LLMs, we evaluated in this work the performance of LLMs in drafting DMPs compliant with the NIH 2023 guidelines.

## Methods

### About the NIH DMP

The 2023 NIH DMP is divided into six main sections, referred to as elements: 1) Data Type, 2) Related Tools/Software/Code, 3) Standards, 4) Data Preservation, Access, and Associated Timelines, 5) Access, Distribution, or Reuse Considerations, and 6) Oversight of Data Management and Sharing. Some of these elements are further divided into subsections, resulting in a total of 12 sub-elements in the NIH DMP. A template, along with 26 DMP samples from actual grant proposals, is available on the NIH "Writing a Data Management & Sharing Plan" webpage (as of June 2025). We used these samples as our reference benchmark to evaluate how closely LLM-generated DMPs align with NIH expectations.

### LLMs Evaluated

In this study, we evaluated the performance of two LLMs: the open-source Llama 3.3 70B model and OpenAI's commercial GPT-4.1.[20,21] Prior studies evaluating LLMs on similar technical writing tasks, including grant writing and clinical document drafting, have found GPT-family models to outperform Llama-family models, while still finding Llama competitive in some contexts.[22–25] We included both here to investigate how an open-source LLM compares with a commercial one in drafting DMPs. We selected Llama 3.3 and GPT 4.1 as the latest models from these families available at the time of our study.

### Prompting Strategy

To identify the information required in an LLM prompt to generate an accurate NIH DMP, we reviewed the NIH-provided DMP template and 26 DMP samples. Because each of the 12 sub-elements requires several pieces of information, we broke them down into what we called core elements, i.e., distinct pieces of information expected in each sub-element. For example, in Element 1B ("Scientific data that will be preserved and shared, and the rationale for doing so: Describe which scientific data from the project will be preserved and shared and provide the rationale for this decision"), we identified four core elements: 1) Type of data that will be preserved and shared, 2) Rationale for sharing the data, 3) Type of data that will not be shared, 4) Rationale for not sharing the data. Overall, we identified 29 unique core elements across the NIH DMP. A full breakdown is provided in Table 1. From these core elements, we identified three that must be included in a prompt for generating a DMP because they are specific to a proposal and cannot be inferred by an LLM: 1) The data types to be collected, 2) The subject type (human or non-human) and the number of subjects anticipated (if applicable), and 3) The consent type (if human subject). In addition, based on the NIH DMP samples, we observed that the NIH Institute/Center where the proposal is submitted also affects aspects of the DMP, such as the repository used for data sharing. Therefore, we decided that the NIH institute/center's name should also be included in the prompt.

Overall, our prompting strategy consisted of a single prompt that included the name of the NIH Institute/Center to which the proposal was submitted, details about data collection (from human or non-human participants, number of subjects, and data types to be collected), and the full NIH-provided DMP template. Note that these are not necessarily the only core elements required to write an accurate DMP, but we limited our selection to these to test a minimal prompt approach.

## Custom LLM Evaluation Framework

### Overall approach

Traditional machine learning models are commonly evaluated using established metrics such as precision, recall, or area under the curve (AUC).[26] However, evaluating LLMs is more challenging because their outputs are open-ended and variable across runs.[27] Most existing LLM evaluation methods assess performance along dimensions such as accuracy, compliance, fluency, coherence, relevance, factual consistency, fairness, safety, robustness, and explainability.[28,29,30,31] There are various methods for evaluating these performance dimensions, including reference-based scoring, reference-free scoring, LLM-based judging, and human expert reviews.[32] Since there are no established standards to evaluate LLM-generated DMPs, we decided to follow an evaluation framework that combines two complementary components: 1) Automatic reference-based evaluation aimed at providing a scalable quantitative comparison across a large set of DMPs and 2) Human expert review aimed at capturing qualitative judgment from experts to assess real-world usability and compliance.

### Automatic reference-based evaluation

For the automatic reference-based evaluation, we used all 26 NIH-provided DMP samples. These samples include their associated NIH Institute/Center but do not include the full proposal details. To provide proposal-specific content to the LLMs, we extracted the content of Element 1A to fill out the study details item in our prompt, since Element 1A is where researchers have to provide an overview of the data they will generate. We also extracted the consent type from Element 5A of these samples if they were related to human studies. The exact prompt structure was as follows:

> "*You are an expert biomedical grant writer and data steward. Create a DMSP for a grant proposal being submitted to the NIH, specifically targeting the [name and acronym of specific NIH institute/Center].*
>
> *[If human study:] This proposal includes a study that will involve human participants. [description of the data sharing consent extracted from Element 5A].*
>
> *Here are the details about the data to be collected: [description of the details about the data to be collected from Element 1A].*
>
> *Follow exactly this markdown format template provided by the NIH to produce a well-formatted markdown output that follows the NIH-required format: [insert NIH template]."*

We ran this prompt 10 times per DMP sample for both the Llama and GPT models to account for variability in LLM outputs.[33] We generated all outputs using a Jupyter Notebook (see

Code Availability section). Llama outputs were generated using a local deployment of Llama 70B 3.3 with Ollama on a workstation with dual RTX 3090 GPUs (24GB VRAM each). For the GPT model, we used the ChatGPT API. All DMPs (NIH, Llama-generated, and GPT-generated) are included in the dataset shared with this work (see the Data Availability section).

To compare the LLM-generated DMPs with the NIH samples, we computed semantic and textual similarity using ROUGE (Recall-Oriented Understudy for Gisting Evaluation) and SBERT (Sentence Bidirectional Encoder Representations from Transformers).[34,35] While ROUGE measures lexical overlap between generated and reference text by comparing n-gram matches, SBERT similarity measures the cosine similarity of embeddings. We evaluated these two metrics at three levels of granularity: 1) the full DMP, 2) the 6-element levels, and 3) the 12 sub-element levels. Evaluating these multiple levels enabled us to assess not only overall similarity but also whether models captured the expected content in specific DMP sections. For each evaluation, we ensured that the template text was removed and that only the filled-out text was compared.

**Human expert evaluation**

Our human expert evaluation consisted of asking each participant to review three randomly assigned and independent DMPs, each from a different sample: 1) An original NIH-provided DMP, 2) A Llama-generated DMP, and 3) A GPT-generated DMP. Because evaluating the full set of 26 NIH sample DMPs would have required more participants than we anticipated recruiting, we selected a smaller, yet representative, sample of 10 DMPs. The complete corpus of 26 NIH DMP samples spans 8 NIH institutes/Centers and 21 tags (keywords provided along with the samples). To construct a representative subset, we implemented a greedy selection algorithm that maximizes coverage across both dimensions. The final selection of 10 DMPs captured all 8 institutes and 21 tags. Of the 10 selected DMPs, 8 involved human data and 2 involved non-human data. The subset also reflected a wide range of data types, including imaging, clinical data, genomics, behavioral research, and qualitative analysis. The 10 selected samples, along with their Llama- and GPT-generated versions (we randomly selected one of the 10 generated for the automatic evaluation), are included in the datasets associated with this work (see the Data Availability section).

We recruited human experts through the panel of user research participants from the DMP Tool, which is primarily made up of data librarians at universities. Invitations to this user research panel were shared on DMP Tool social media sites, the DMP Tool blog, conference presentations about DMPs, and in relevant academic librarian forums. Participation was fully anonymous and did not include any identifying information. We did not collect personal data such as names, email addresses, institutional IDs, IP addresses, or any other identifiers. The University of California, Davis reviewed the protocol and determined that it did not meet the DHHS or FDA definitions of human subjects research, so IRB review was not required (IRB ID 2334657-1; June 24, 2025).

To conduct the evaluation, we built a custom survey website because existing survey platforms did not meet our needs. Participants accessed the survey via a protected link sent in the invitation. The full evaluation survey is included in the dataset of this work (see Data Availability section). In the first part of the survey, participants were asked to provide information about their professional role, experience with DMPs, and attitudes toward the use of AI in research. All background questions were optional, but they were provided by all the participants and used to contextualize their feedback when provided. In the second part of the survey, they were asked to evaluate their three randomly assigned DMP samples one by one. For each DMP, they were asked to download the PDF of the DMP, read it, and then evaluate each of the 6 elements one by one through three questions per element (each element was also shown on the survey screen during rating to support evaluation):

- Satisfaction: 1)"How satisfied are you with the response to this element?" on a five-point scale ranging from 1 ("Not at all satisfied") to 5 ("Extremely satisfied")
- Error types (select all that apply): "What type of errors did you find, if any?" with nine options, including
  - Accuracy (incorrect or misleading information)
  - Completeness (missing required content for the element)
  - Clarity (vague language)
  - Coherence (information not logically organized or well connected)
  - Compliance (the DMP would not comply with funder policy)
  - Fluency (spelling, punctuation, or grammar issues)
  - Best Practice (choices that are not recommended)
  - Other (free-text comment)
  - None (no errors observed).
- Comments: Participants were invited to provide additional comments to highlight issues not captured by the structured questions.

After completing the element-wise evaluation, participants were asked to provide an overall evaluation of each DMP version by answering two questions: 1) How satisfied they were with the DMP as a whole (1 to 5) and 2) Whether they believed the DMP was generated by a human or an LLM. All survey responses were stored in a structured JSON format to support downstream processing.

We used Python scripts to analyze the responses using libraries such as pandas[36] and Numpy[37] for data analysis, Scipy[38] for statistical analysis, and Matplotlib[39] and seaborn[40] for data visualization (see Code Availability section). We averaged their overall satisfaction ratings for each DMP version and their per-element satisfaction ratings. To analyze their error ratings, we summed the counts of error types across the six elements to obtain the total count per error type for each DMP version. We then summed the error counts for each type across all evaluations and divided by the total to compute the mean number of errors per DMP per error type.

We assessed whether participant satisfaction differed across the three DMP versions by treating each participant as a paired observation and applying the Friedman test to the mean satisfaction across the six NIH elements, both overall and for each element. Where the Friedman test was significant, we conducted post hoc pairwise comparisons using Wilcoxon signed-rank tests with Holm correction for multiple comparisons.[41,42] As a sensitivity analysis that uses every element-level rating rather than per-participant means, we additionally fitted a cumulative link mixed model[43] and a linear mixed model[43], each with a random intercept for participant and a fixed effect for the DMP element. The Friedman and Wilcoxon tests were computed in SciPy, the ordinal model in the R ordinal package, and the linear mixed model in Python.

## Results

### Automatic Evaluation

The automatic reference-based evaluation showed that GPT achieved higher similarity to the NIH reference DMPs than Llama in both semantic similarity (SBERT) and lexical overlap (ROUGE) (Table 2). SBERT scores range from 0 to 1, where scores above 0.5 generally indicate meaningful semantic similarity and scores around 0.8 or higher reflect strong alignment. ROUGE scores also range from 0 to 1, where scores above 0.3 are generally considered moderate overlap and scores above 0.5 reflect strong lexical similarity. Averaged across all samples (each score computed as the mean over 10 generations), GPT scored slightly higher on SBERT (0.64) than Llama (0.61) and slightly higher on ROUGE (0.28) than Llama (0.26). Performances also varied across samples (Table 2). Treating each DMP as a paired observation (n=26), GPT's higher performance over Llama was statistically significant for both SBERT (mean $\Delta$=+0.03, Confidence Interval CI [+0.02, +0.04], paired t-test $p = 6.6 \times 10^{-8}$, Wilcoxon signed-rank $p = 4.2 \times 10^{-7}$) and ROUGE (mean $\Delta$=+0.018, Confidence Interval CI [+0.010, +0.027], paired t-test $p = 9.7 \times 10^{-5}$, Wilcoxon signed-rank $p = 2.5 \times 10^{-4}$). GPT scored higher in 24/26 DMPs for SBERT and 20/26 (3 ties) for ROUGE. The DMP-wise breakdown of the evaluation is included in the dataset (see Data Availability section).

Element-level automatic evaluation showed that GPT generally matched the NIH reference DMPs more closely than Llama, with the clearest improvements in policy and infrastructure-specific Elements (Table 3). Element 1 showed the highest match, as expected, since the content from Element 1A of the NIH reference DMPs was included in the prompts. Sub-element-level analysis provided a more granular view of where models aligned with the NIH reference DMPs (Table 4). GPT showed its largest gains over Llama in Elements 1B (Data to be preserved and shared), 4A (Repository), and 5A (Access, Distribution, Reuse), suggesting that GPT better handles content requiring specific technical and policy knowledge.

### Human Expert Evaluation

A total of 21 participants completed the human expert evaluation (Fig. 1). Participants were primarily from Library and Information Science (54.5%) and Biological and Life Science (31.7%). Most participants identified their current role as librarian/data librarian/research support staff (57.1%), followed by data managers/stewards (21.4%). Participants had substantial experience in their roles, with the largest group reporting 4–6 years (42.9%), and many reporting more than 7 years (33.3%). They reported reviewing DMPs relatively frequently (most commonly about once a month; 47.6%) and were generally familiar with NIH DMPs (moderately familiar: 33.3%; extremely familiar: 42.9%). A majority reported not using AI tools like ChatGPT in their work (57.1%), and most reported being slightly (47.6%) or somewhat (19.0%) comfortable using AI to write DMPs. Their overall attitudes towards the use of AI in research workflows leaned cautiously positive, with 42.9% reporting being somewhat or very in favor, compared with 33.4% somewhat or very against. The full summary table of the background responses is included in the dataset (see Data Availability section).

Across the 21 participants, the three versions (human, GPT, Llama) of the 10 DMP samples retained were each evaluated 2 to 3 times. Overall satisfaction ratings varied substantially across the three versions (Fig. 2). GPT received predominantly positive ratings, with 19/21 (90.5%) satisfied or very satisfied. In contrast, Llama received substantially lower satisfaction ratings, with only 4/21 (19.0%) reporting satisfaction or very satisfied. The human-written, NIH-provided DMPs showed a more mixed distribution, with 9/21 (42.9%) satisfied or very satisfied.

The Element-level satisfaction ratings were consistent with the overall ratings (Table 5). On a 1-5 satisfaction scale, GPT received the highest mean ratings for every element (ranging from 3.9 to 4.4), with particularly strong scores for Elements related to Tools/Software/Code, Standards, and Oversight (Fig. 5). In contrast, Llama received lower ratings overall (range 2.5-3.4), with the lowest scores on Elements related to Repository/Findability/Timeline and Access/Distribution/Reuse and Privacy. Human-written DMPs were generally rated between the two models (range 2.9-3.4). They were rated higher than Llama on Elements 4, 5, and 6, but lower than GPT on all elements (Table 5).

These differences were statistically significant. Treating each participant as a paired observation (n = 21), overall satisfaction differed across the three versions (Friedman test, $\chi^2 = 15.41$, $p = 4.5 \times 10^{-4}$). Post-hoc Wilcoxon signed-rank tests with Holm correction showed that GPT was rated significantly higher than Llama ($p = 5.7 \times 10^{-5}$) and than the human-written DMPs ($p = 0.005$), whereas the human and Llama versions did not differ ($p = 0.701$). This pattern was consistent across all six elements, where the Friedman test was significant in every case (Table 6). Effect sizes for overall participant satisfaction estimated from two mixed models fitted to all element-level ratings (377 observations from 21 participants), each with a random intercept for participant and an adjustment for the DMP element. The cumulative link mixed model (CLMM) treats satisfaction as ordinal and reports proportional-odds ratios, where values above 1 indicate higher satisfaction for the first-named version. The linear mixed model (LMM) treats satisfaction as continuous and reports Cohen's d. Both models agree with the Friedman and Wilcoxon results in Table 6. An ordinal mixed model fitted to all element-level ratings agreed with these results, with GPT showing about nine times higher odds of a more satisfied rating than Llama (odds ratio 9.1) and about seven times higher odds than the human version (odds ratio 7.0). A linear mixed model gave correspondingly large effect sizes (Cohen's d of 1.06 and 0.93; Table 7).

Error analysis (Fig. 3) further reinforced these patterns. GPT consistently produced the fewest errors, according to the evaluators (48/21 total errors across all evaluators and error types), with particularly low error rates for coherence (0.09 errors per DMP), compliance (0.19), fluency (0.19), and completeness (0.23). Human-authored DMPs fell in between GPT and Llama overall (147/21 total errors), with evaluators reporting high error rates for clarity (1.52) and completeness (1.52). Llama received the highest error counts from evaluators (153/21 total errors), with particularly high error rates for clarity (2.52) and completeness (1.61). Across all three versions, clarity and completeness were the most frequently flagged error categories by evaluators, suggesting that producing clearly written, complete DMPs remains a key challenge regardless of who writes them.

Finally, when asked to judge whether each of the three DMP versions they were evaluating was human-written or LLM-generated (Fig. 4), GPT outputs were most often mistaken for human-written text, with only 3/21 (14.3%) correctly identifying them as LLM-generated. In contrast, Llama outputs were usually recognized as LLM-generated, with 15/21 (71.4%) correct identifications. Human-written DMPs were generally recognized as human, with 14/21 (66.7%) correct classifications. These results indicate that GPT DMPs were substantially more “human-like” to participants than Llama outputs.

### Synthesis of Findings

Across both evaluation components, GPT outperformed Llama and, in some respects, human-written DMPs. The automatic evaluation showed that both models achieved moderate semantic alignment with NIH-provided sample DMPs and relatively low lexical alignment, as expected given the open-ended nature of DMPs. GPT’s better performance over Llama, while statistically significant, was modest in absolute terms. The human expert evaluation revealed a more pronounced difference between the models, as GPT-generated DMPs received substantially higher satisfaction ratings, fewer flagged errors, and were nearly indistinguishable from human writing according to the evaluators. However, even GPT-generated DMPs recorded a non-negligible number of errors flagged by evaluators (48 across 21 GPT-generated DMPs), particularly regarding clarity and completeness. These findings suggest that while LLM-assisted DMP drafting is promising, human review remains an important step.

## Discussion

In this work, we evaluated the performance of two LLMs, GPT-4.1 and Llama 3.3, in drafting DMPs compliant with the NIH 2023 guidelines using simple research inputs. This is intended as an initial step toward understanding the broader potential of LLMs to support researchers and librarians in this challenging task. Across both automatic and human evaluations, GPT consistently outperformed Llama. GPT-draft achieved higher semantic and lexical similarity to NIH reference DMPs than Llama did, particularly for elements and sub-elements that require specific technical and policy knowledge, such as repository selection, access conditions, reuse considerations, and oversight. Blinded human expert evaluation also aligned with these results, as GPT drafts received the highest satisfaction ratings and the fewest reported errors, while Llama drafts were rated lowest and had the highest error counts. Human-written, NIH-provided samples showed a more mixed profile overall, generally falling between GPT and Llama in perceived quality.

A particularly striking finding was that evaluators frequently misclassified GPT outputs as human-written, whereas they correctly identified Llama outputs as LLM-generated most of the time. Interestingly, human-written DMPs were misclassified as LLM-generated in roughly one-third of the cases. This suggests that GPT-generated DMPs not only match the expected tone and structure of NIH DMPs but may, in some respects, exceed the quality of DMPs produced by researchers and librarians. Overall, these results suggest that LLMs, particularly commercial models like GPT, could reduce the burden of preparing DMPs for researchers and librarians lacking training in FAIR data practices, as well as for institutions with limited resources dedicated to data management. This observation still comes with important caveats, since even GPT-generated DMPs were not error-free, according to the evaluators, who reported errors particularly in clarity and completeness. This underscores that LLMs should be positioned as drafting aids rather than autonomous solutions.

An important methodological observation from our study is the divergence between automatic and human evaluations in the magnitude of the differences detected between the two models. The automatic evaluation showed that GPT outperformed Llama with statistical significance, but the absolute differences were small, suggesting that the models were broadly similar when measured by textual and semantic similarity alone. The human expert evaluation, however, revealed a far more pronounced gap in favor of GPT across satisfaction ratings, errors flagged, and human-like writing. Statistical testing confirmed that GPT's higher performance in human satisfaction was significant across all six elements. This consistency suggests the advantage reflects general drafting quality rather than strength in a particular part of the plan. These data highlight an important limitation of automatic similarity metrics for evaluating open-ended documents like DMPs: they are scalable and easier to apply to large samples, but may not capture real-world performance across the qualities that expert evaluators care about.

This study has several limitations. First, the number of reviewers was modest, and each reviewed only one DMP sample of each generation type. Second, our prompting strategy used a single standardized prompt and relied on Element 1A text from the NIH samples to provide context for the proposal, as full proposal narratives were unavailable. This likely limited how tailored the generated plans could be, especially in sections governed by domain norms. Third, our automatic evaluation relied on ROUGE and SBERT, which measure textual and semantic similarity rather than compliance or practical adequacy. Despite these limitations, the overall trends were consistent across automatic scoring and blinded human expert review, supporting the robustness of our main conclusions.

Another caveat of this work is that creating a good quality DMP on its own is not the only important metric for whether LLMs can assist in drafting DMPs. It is important to also ensure that researchers understand and accept the plan as one they can commit to, rather than just copying the text of the LLM-generated DMP uncritically. The first step is making sure that LLMs can generate a good plan, which this study investigated, while future work on tooling and implementing can identify the best ways to ensure researchers are engaged in the process and making important decisions with the guidance of the LLM assistant.

Since the completion of this study, the NIH has released updated DMP guidelines that will be enforced starting from the end of May 2026. The new guidelines include a simplified DMP template with mostly yes/no questions and some details about the data to be collected. While this means that the specific DMP template evaluated here is no longer the current NIH standard, we believe this does not diminish the relevance of our findings since our goal was not to evaluate LLM performance on a particular template, but to assess the broader potential of LLMs to support DMP drafting. We believe that our findings remain pertinent regardless of the required DMP format. If anything, a simpler template with more constrained response formats may further lower the bar for LLM-assisted drafting, suggesting that the promises identified in this work would likely extend to the new NIH DMP template. Moreover, the evaluation framework we developed is template-agnostic and can be readily applied to assess LLM performance on any DMP template. We therefore view this study as establishing a methodological foundation and an initial benchmark rather than a definitive evaluation tied to a single template.

Our future work will extend this evaluation to other major funders, including NSF, and to the updated NIH guidelines, to assess whether the findings generalize across different DMP formats and requirements. We will also investigate how enhancement approaches, such as retrieval-augmented generation (RAG), can improve the performance of open-source LLMs, including Llama, to make high-quality LLM-assisted DMP drafting accessible to research institutions that cannot rely on commercial APIs.

## Data Availability

The data associated with this manuscript consists of DMP samples (Human, GPT, and Llama-generated versions), the full evaluation survey dataset, the DMP-level evaluation breakdown, the complete summary of background survey responses, and full error counts organized by NIH DMP elements and model type. Since no FAIR guidelines were found for structuring this data, we used the SPARC Data Structure (SDS), which provides a broad data and metadata framework for organizing research data in line with the FAIR principles.[44] The SPARC data curation software SODA for SPARC was used to organize the data and prepare the metadata files.[45,46] The dataset is maintained in a GitHub repository called "nih-dmp-llm-evaluation-paper-data" in the FAIR Data Innovations Hub GitHub organization, and the version associated with this manuscript (v1.0.0) is also archived on Zenodo.[47] This data is shared under the permissible Creative Commons Attribution 4.0 International (CC-BY) license.

## Code Availability

The code associated with this manuscript is maintained across two GitHub repositories in the fairdataihub GitHub organization. One repository, called "nih-dmp-llm-generation", contains the notebook and instructions for generating DMPs with Llama and GPT. The other repository, called "nih-dmp-llm-evaluation-paper-code" contains two notebooks related to our automated and human evaluations. The code was made FAIR in accordance with the FAIR-BioRS guidelines using Codefair.[48,49] Accordingly, the versions of the code associated with this manuscript (v1.0.0) are archived on Zenodo.[50,51] The notebook and associated files are shared under the permissive MIT license.

## Acknowledgement

We thank the domain-expert reviewers for their time and feedback, and we acknowledge the project teams for their operational and technical support.

## Author Contributions

N.Z., R.H.G., M.P., J.C., and B.P. conceived the study. N.Z., J.O., X.D., R.H.G., and B.P. designed the methodology. N.Z., J.O., and X.D. developed the software. N.Z. and B.P. curated the data and wrote the original manuscript draft. N.Z., J.O., X.D., R.H.G., B.R., and B.P. carried out the investigation. N.Z., J.O., and B.P. performed the formal analysis. N.Z., J.O., B.R., and B.P. validated the results. N.Z. and J.O. prepared the figures. M.P., J.C., and B.P. provided resources, supervised the work, and acquired funding. R.H.G., M.P., and J.C. administered the project. All authors reviewed and approved the final manuscript.

## Competing interests

The authors declare no competing interests.

## Funding

This work was supported by the National Science Foundation (NSF) and the Chan Zuckerberg Initiative (CZI), supported in part by NSF Award No. 2332353.

# Figures

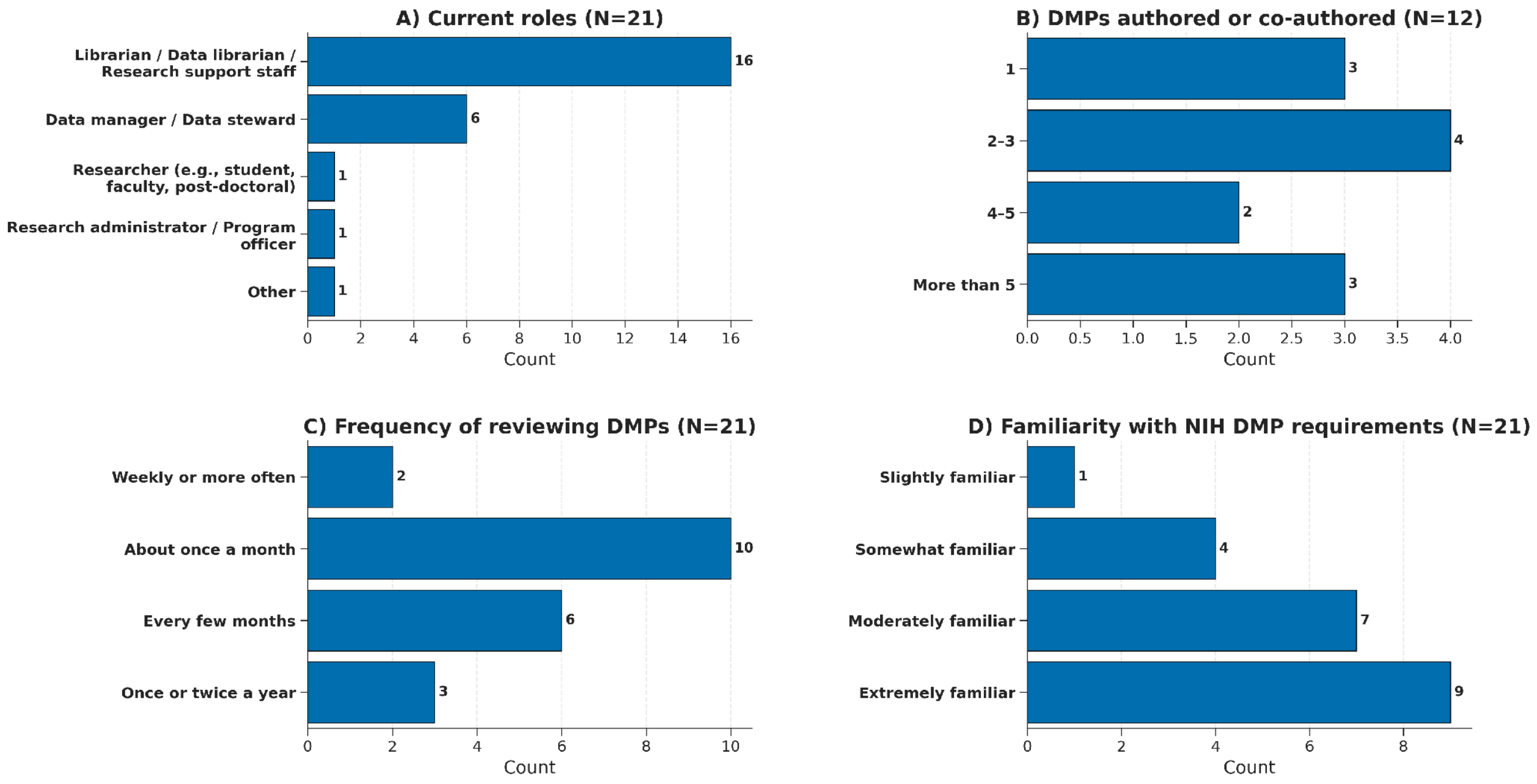

**Figure 1**. Background and DMP expertise of participants in our human expert evaluation. A) Current roles (multi-selection allowed). B) Number of DMP authors or co-authors. C) Frequency of reviewing DMPs. D) Familiarity with NIH DMP requirements. Responses to all Background survey items were optional.

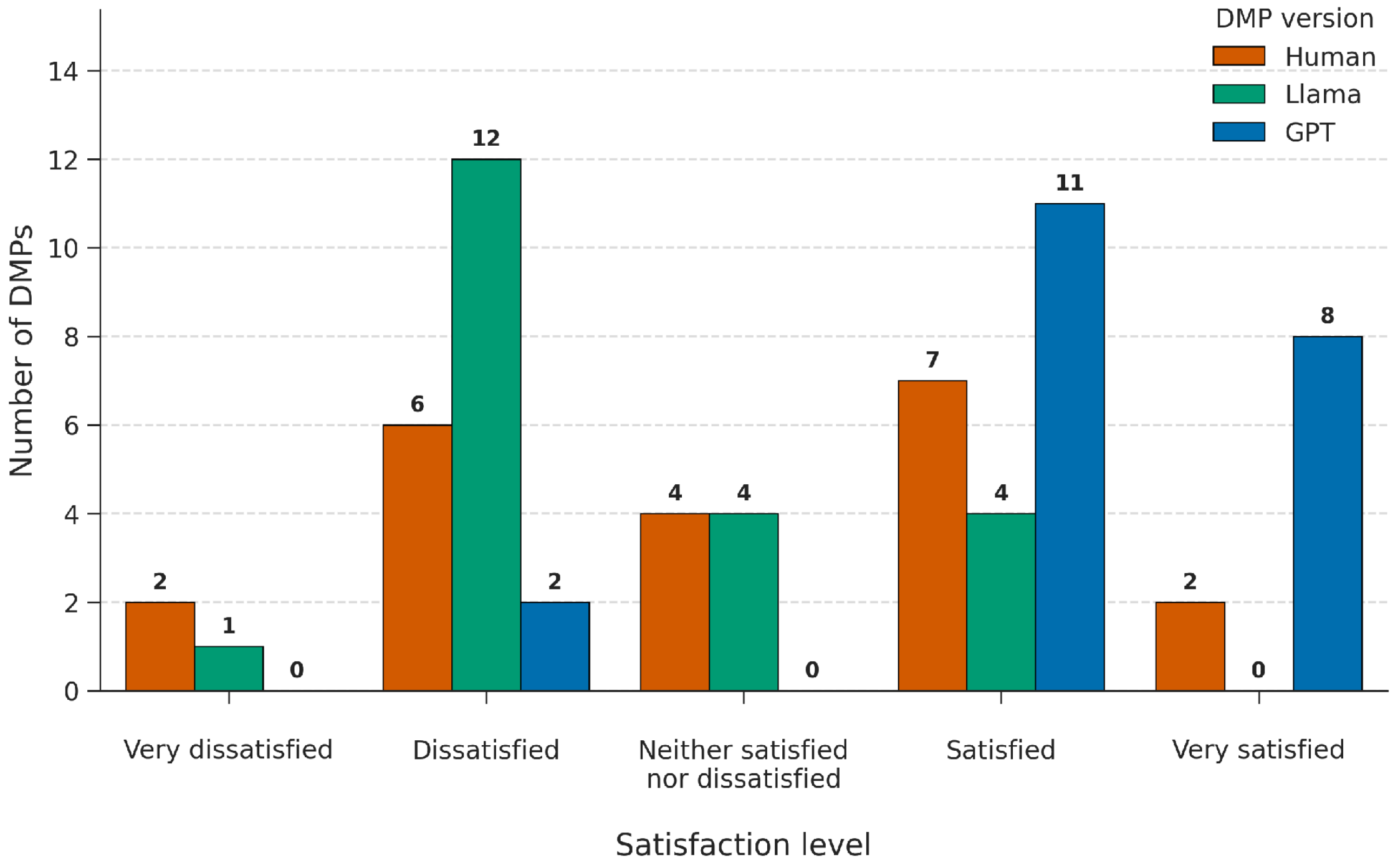

**Figure 2**. Distribution of overall satisfaction of the participants for the different DMP versions.

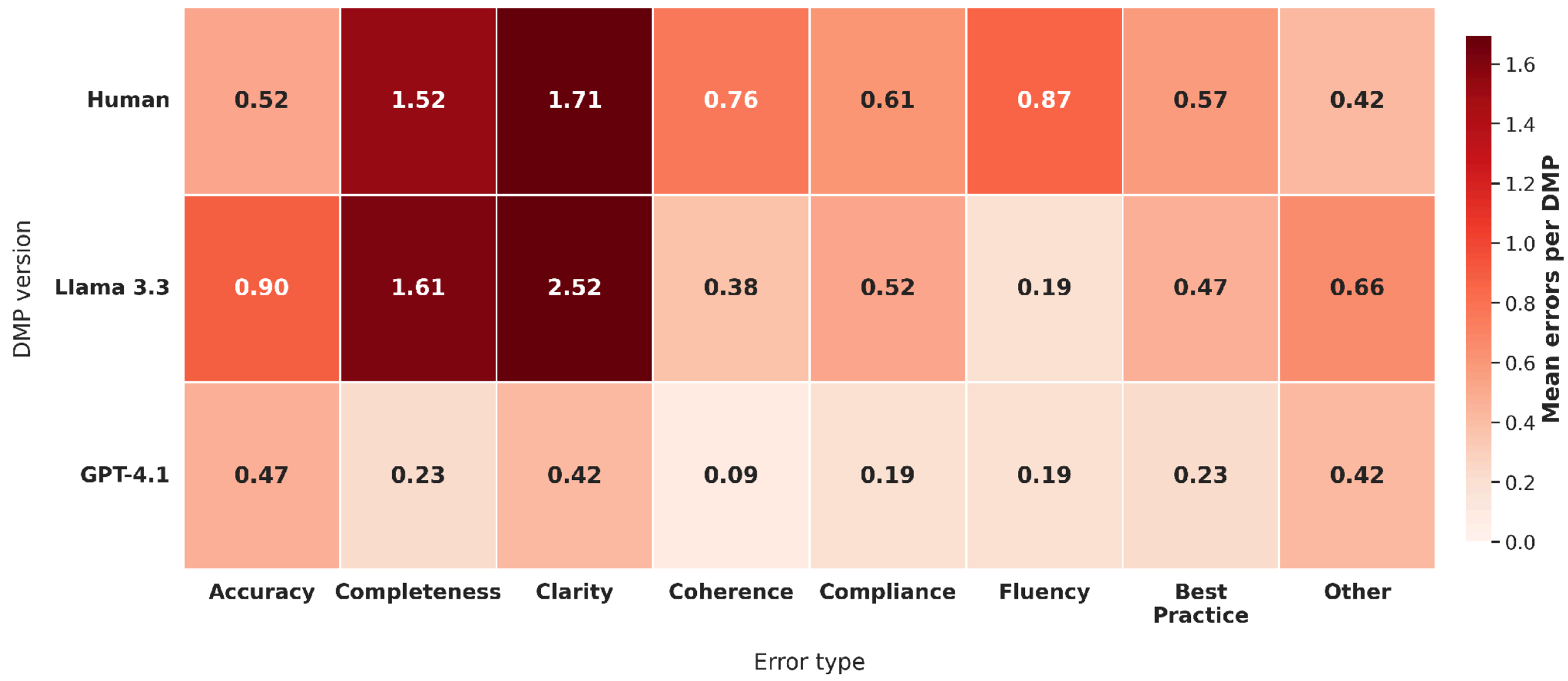

**Figure 3**. Average number of errors per DMP as assessed by human evaluators across eight error types and the three DMP versions (N=21 each) included in the human evaluation.

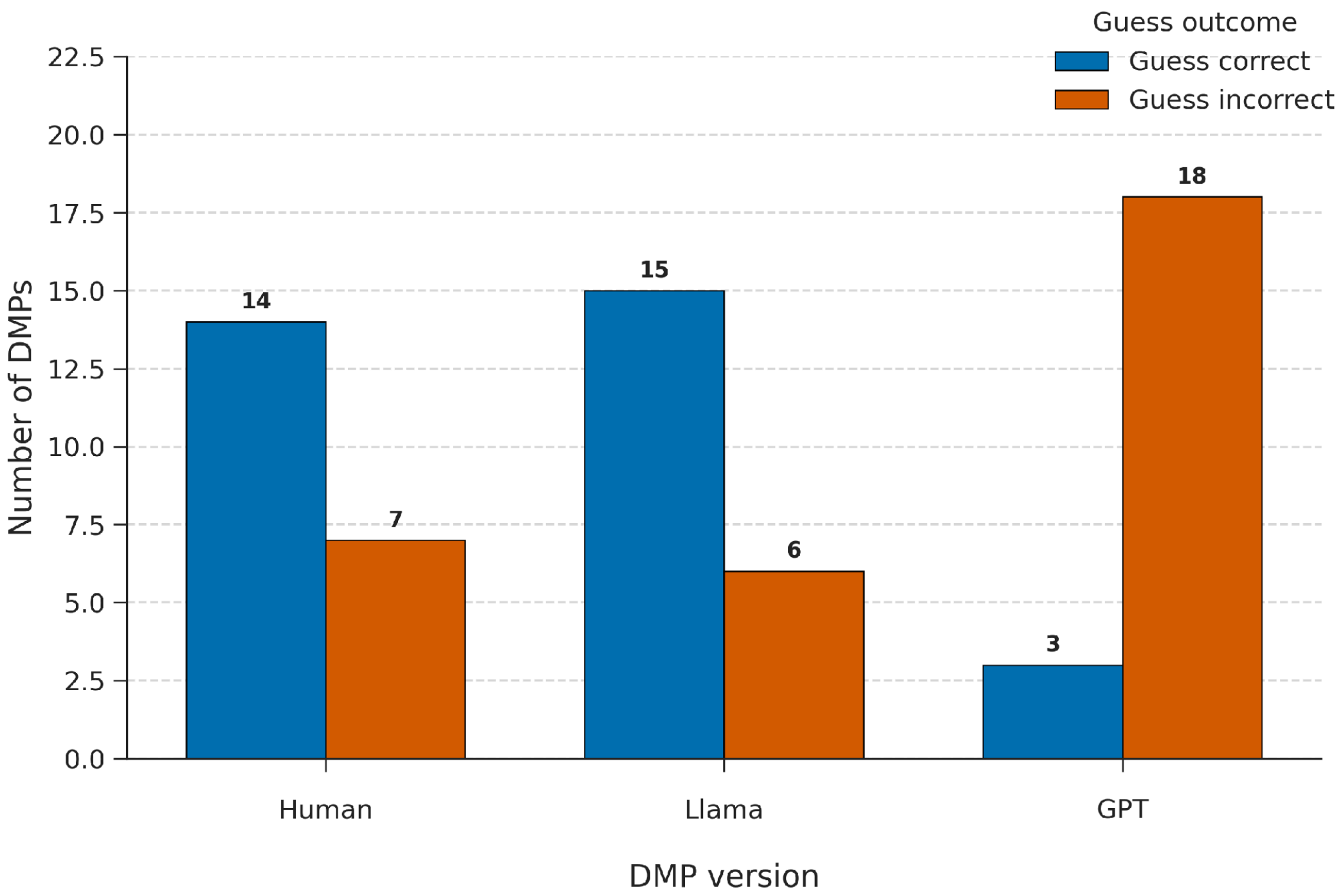

**Figure 4**. Outcomes of the participants' responses when asked to guess if a given DMP version was human-written or LLM-generated.

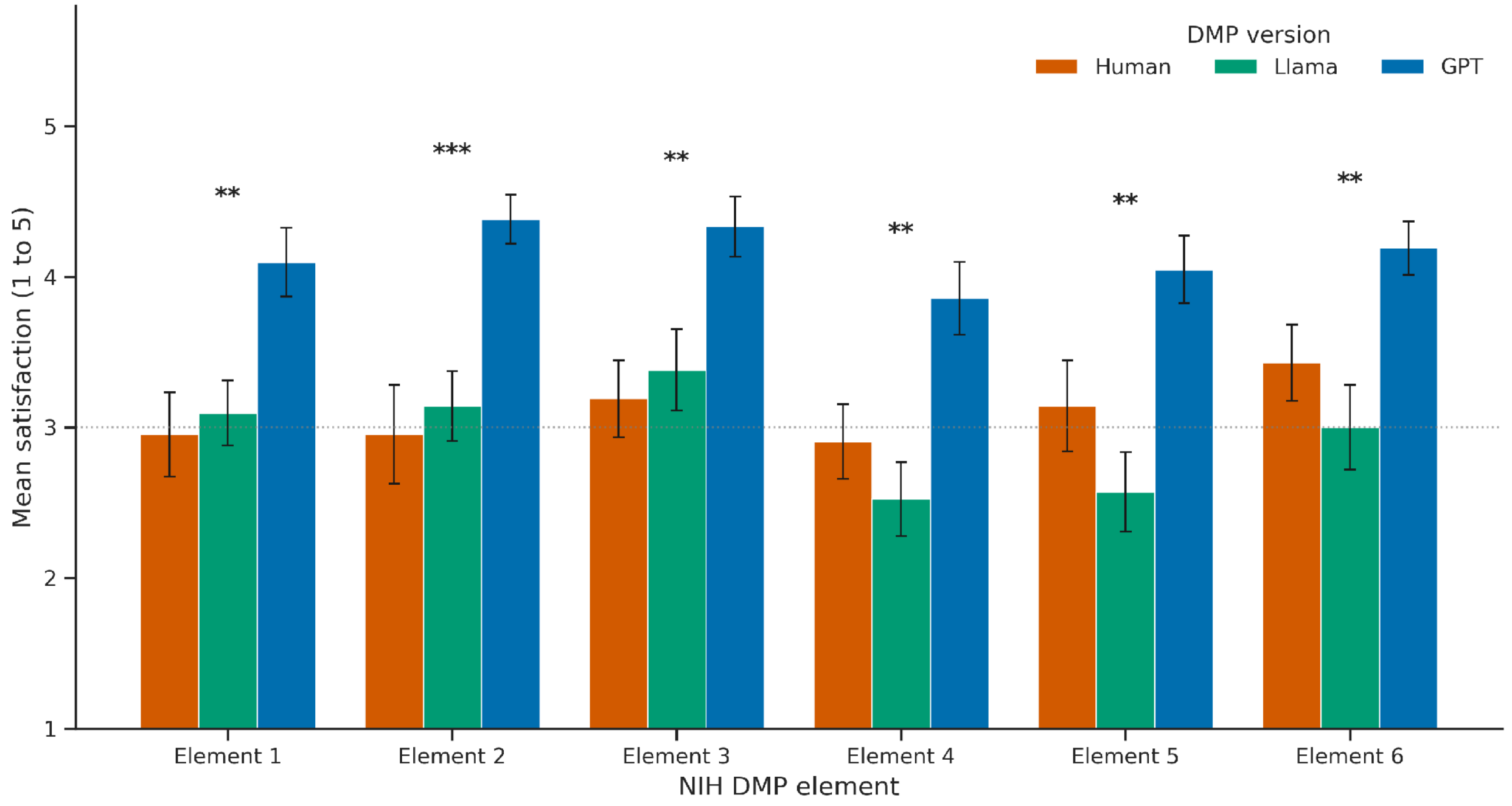

**Figure 5.** Mean participant satisfaction with each NIH DMP element by DMP version (Human, Llama 3.3, GPT 4.1). Bars show the mean satisfaction on a 1 to 5 scale, and error bars show the standard error of the mean. The marker above each element reports the Friedman test for that element (* $p < 0.05$, ** $p < 0.01$, *** $p < 0.001$), which tests whether the three versions differ; the pairwise comparisons are given in Table 6. The dotted line marks the neutral midpoint of the scale.

## Tables

**Table 1**. Breakdown of the 12 sub-elements of the NIH DMP template into core elements.

| NIH DMP sub-element | Core elements |
|---|---|
| 1. A. Types and Amount of Scientific Data | Data types to be collected; Amount of data expected to be generated; Subject type (human or non-human), and the number of subjects anticipated (if applicable) |
| 1. B. Data to be Preserved and Shared | Type of data that will be preserved and shared; Rationale for sharing the data; Type of data that will not be shared; Rationale for not sharing the data |
| 1. C. Metadata and Documentation | Metadata, documentation, and other related resources that will be shared, besides the data |
| 2. Tools, Software, Code | Name and description of the tools, software, and/or code required to manipulate each data type; Links/references of the tools, software, and/or code required to manipulate each data type; Indication of whether the tools, software, and/or code are open source and free or not; Indication of how the tools, software, and/or code can be accessed |
| 3. Standards | File format that will be used for each data type; File format to be used for the metadata that will be provided with the data; Name of the standards that will be followed for making the data and metadata interoperable; Explanation of how the standards will be applied; Statement that no standard exists for a given data type (if applicable) |
| 4. A.Repository | Name of the repository(ies) where scientific data and metadata will be shared |
| 4. B. Findability | Description of how data will be findable and identifiable (e.g., via a persistent and unique identifier) |
| 4. C. Availability Timeline | When the data will be made available to others (must be by the time an associated publication is made or before the end of the grant performance period); How long the data will be available |
| 5. A. Access, Distribution, Reuse | Consent type (if human subject); Explanation of privacy or confidentiality protections that will be applied; Explanation of any data use agreements or restrictions on redistribution |
| 5. B. Controlled Access | Whether access to data will be controlled; Explanation of the data access process if controlled |
| 5. C. Privacy and Confidentiality | Describe how the privacy, rights, and confidentiality of human research participants will be protected, e.g., through de-identification, Certificates of Confidentiality, and other protective measures (if applicable) |
| 6. Oversight of DMP | Indication of who will monitor and manage compliance with this plan; Frequency of monitoring compliance with this plan |

**Table 2**. Aggregate summary of automatic reference-based evaluation at the full DMP level across 26 NIH DMP samples (10 LLM generations per sample per model).

| Summary | Llama 3.3 | GPT 4.1 | Δ(GPT-Llama) |
|---|---|---|---|
| SBERT means across DMPs | 0.61 | 0.64 | +0.03 |
| SBERT range across DMPs (min-max) | 0.53 - 0.67 | 0.55 - 0.71 | +0.02 / +0.04 |
| # DMPs where the average SBERT is higher in GPT | | 24/26 | |
| ROUGE mean across DMPs | 0.26 | 0.28 | +0.02 |
| ROUGE range across DMPs (min-max) | 0.18 - 0.36 | 0.19 - 0.38 | +0.19 / +0.38 |
| # DMPs where the average ROUGE is higher in GPT | | 20/26 (3 ties) | |

**Table 3**. Aggregate summary of automatic reference-based evaluation at the element level. The mean and SD are computed across the 260 generated element responses (10 generations per sample) for each model.

| Element # + information asked | SBERT Llama 3.3 | SBERT GPT 4.1 | Δ SBERT (GPT-Llama) | ROUGE Llama 3.3 | ROUGE GPT 4.1 | Δ ROUGE (GPT-Llama) |
|---|---|---|---|---|---|---|
| 1. Types and amount of scientific data, Data to be preserved and shared; Metadata and documentation | 0.82 ± 0.08 | 0.82 ± 0.07 | 0.00 | 0.37 ± 0.12 | 0.34 ± 0.09 | -0.03 |
| 2. Tools, software, and code | 0.61 ± 0.10 | 0.59 ± 0.11 | -0.02 | 0.19 ± 0.09 | 0.24 ± 0.10 | +0.05 |
| 3. Standards | 0.57 ± 0.14 | 0.60 ± 0.13 | +0.03 | 0.21 ± 0.09 | 0.24 ± 0.10 | +0.03 |
| 4. Repository, Findability, and Availability timeline | 0.65 ± 0.09 | 0.69 ± 0.08 | +0.04 | 0.25 ± 0.11 | 0.26 ± 0.11 | +0.01 |
| 5. Access, Distribution, Reuse, Controlled Access, Privacy, and Confidentiality | 0.71 ± 0.09 | 0.75 ± 0.08 | +0.04 | 0.21 ± 0.09 | 0.24 ± 0.09 | +0.03 |
| 6. Oversight of DMP | 0.63 ± 0.14 | 0.65 ± 0.11 | +0.02 | 0.22 ± 0.09 | 0.25 ± 0.09 | +0.03 |

**Table 4**. Aggregate summary of automatic reference-based evaluation at the sub-element level. The mean and SD are computed across the 260 generated element responses (10 generations per sample) for each model.

| Sub-element # + information asked | SBERT Llama 3.3 | SBERT GPT 4.1 | Δ SBERT (GPT-Llama) | ROUGE Llama 3.3 | ROUGE GPT 4.1 | Δ ROUGE (GPT-Llama) |
|---|---|---|---|---|---|---|
| 1A. Types and Amount of Scientific Data | 0.87 ± 0.10 | 0.82 ± 0.09 | −0.05 | 0.54 ± 0.22 | 0.44 ± 0.15 | −0.10 |
| 1B. Data to be Preserved and Shared | 0.50 ± 0.15 | 0.60 ± 0.12 | 0.1 | 0.26 ± 0.14 | 0.30 ± 0.13 | 0.04 |
| 1C. Metadata and Documentation | 0.62 ± 0.10 | 0.63 ± 0.10 | 0.01 | 0.22 ± 0.11 | 0.22 ± 0.11 | 0.00 |
| 2. Tools, Software, Code | 0.61 ± 0.10 | 0.59 ± 0.11 | −0.02 | 0.19 ± 0.09 | 0.24 ± 0.10 | 0.05 |
| 3. Standards | 0.57 ± 0.14 | 0.60 ± 0.13 | 0.03 | 0.21 ± 0.09 | 0.24 ± 0.10 | 0.03 |
| 4A. Repository | 0.47 ± 0.14 | 0.53 ± 0.16 | 0.06 | 0.25 ± 0.12 | 0.29 ± 0.14 | 0.04 |
| 4B. Findability | 0.63 ± 0.10 | 0.66 ± 0.09 | 0.03 | 0.26 ± 0.13 | 0.28 ± 0.14 | 0.02 |
| 4C. Availability Timeline | 0.65 ± 0.08 | 0.66 ± 0.07 | 0.01 | 0.27 ± 0.16 | 0.28 ± 0.17 | 0.01 |
| 5A. Access, Distribution, Reuse | 0.58 ± 0.13 | 0.64 ± 0.11 | 0.06 | 0.24 ± 0.20 | 0.30 ± 0.22 | 0.06 |
| 5B. Controlled Access | 0.57 ± 0.09 | 0.61 ± 0.09 | 0.04 | 0.22 ± 0.12 | 0.25 ± 0.12 | 0.03 |
| 5C. Privacy and Confidentiality | 0.60 ± 0.15 | 0.66 ± 0.13 | 0.06 | 0.22 ± 0.13 | 0.25 ± 0.11 | 0.03 |
| 6. Oversight of DMP | 0.63 ± 0.14 | 0.65 ± 0.11 | 0.02 | 0.22 ± 0.09 | 0.25 ± 0.09 | 0.03 |

**Table 5**. Element-level satisfaction rating on a 1-5 scale (1 = Not at all satisfied, 5 = Extremely satisfied). Values are mean ± SD across all 21 evaluations per version (Human, Llama, GPT). Delta represents the difference in the mean values.

| Element number | Human | Llama 3.3 | GPT4.1 | Δ (GPT-Llama) | Δ (GPT-Human) |
|---|---|---|---|---|---|
| 1. Types and amount of scientific data, Data to be preserved and shared; Metadata and documentation | 3.0 ± 1.3 | 3.1 ± 1.0 | 4.1 ± 1.0 | 1.0 | 1.1 |
| 2. Tools, software, and code | 3.0 ± 1.5 | 3.1 ± 1.1 | 4.4 ± 0.7 | 1.3 | 1.4 |
| 3. Standards | 3.2 ± 1.1 | 3.4 ± 1.1 | 4.3 ± 0.9 | 0.9 | 1.1 |
| 4. Repository, Findability, and Availability timeline | 2.9 ± 1.1 | 2.5 ± 1.1 | 3.9 ± 1.1 | 1.4 | 1.0 |
| 5. Access, Distribution, Reuse, Controlled Access, Privacy, and Confidentiality | 3.1 ± 1.4 | 2.6 ± 1.2 | 4.0 ± 1.0 | 1.4 | 0.9 |
| 6. Oversight of DMP | 3.4 ± 1.2 | 3.0 ± 1.3 | 4.2 ± 0.8 | 1.2 | 0.8 |

**Table 6.** Statistical comparison of participant satisfaction across the three DMP versions (Human, Llama 3.3, GPT 4.1), overall and for each of the six NIH DMP elements. For each row, the Friedman test assessed whether satisfaction differed among the three versions, treating each participant as a paired observation. Post hoc pairwise comparisons were performed using Wilcoxon signed-rank tests with Holm correction, and the values shown are the Holm-adjusted p-values. N is the number of participants who rated all three versions; one participant did not rate Element 6 for the Llama version. * p < 0.05, ** p < 0.01, *** p < 0.001; ns, not significant.

| **DMP element** | N | **Friedman $\chi^2$** | **Friedman p** | **GPT vs Llama** | **GPT vs Human** | **Human vs Llama** |
|---|---|---|---|---|---|---|
| Overall (mean across the six elements) | 21 | 15.41 | 4.5 x 10-4 | 5.7 x 10-5 *** | 0.005 ** | 0.701 ns |
| 1. Types and amount of scientific data, Data to be preserved and shared; Metadata and documentation | 21 | 9.51 | 0.009 | 0.028 * | 0.017 * | 0.735 ns |
| 2. Tools, software, and code | 21 | 17.85 | 1.3 x 10-4 | 3.2 x 10-4 *** | 0.002 ** | 0.635 ns |
| 3. Standards | 21 | 13.32 | 0.001 | 0.003 ** | 0.003 ** | 0.685 ns |
| 4. Repository, Findability, and Availability timeline | 21 | 13.30 | 0.001 | 3.2 x 10-4 *** | 0.010 * | 0.244 ns |
| 5. Access, Distribution, Reuse, Controlled Access, Privacy, and Confidentiality | 21 | 12.24 | 0.002 | 0.004 ** | 0.077 ns | 0.246 ns |
| 6. Oversight of DMP | 20 | 9.24 | 0.010 | 0.020 * | 0.136 ns | 0.365 ns |

**Table 7**. Mixed-model effect size estimates for participant satisfaction across DMP versions.

| Comparison | CLMM odds ratio (95% CI) | CLMM *p* | LMM Cohen's d |
|---|---|---|---|
| GPT vs Llama | 9.13 (5.48 to 15.18) | 5.1 x 10-17 | 1.06 |
| GPT vs Human | 6.98 (4.22 to 11.57) | 8.8 x 10-14 | 0.93 |
| Human vs Llama | 1.31 (0.83 to 2.06) | 0.248 | 0.12 |